\documentclass[prl,aps,showpacs,twocolumn,english,unsortedaddress]{revtex4-1}
\pdfoutput=1

\usepackage[T1]{fontenc}
\usepackage[latin9]{inputenc}
\usepackage{babel}
\usepackage{textcomp}
\usepackage{amsmath}
\usepackage{amssymb}
\usepackage{graphicx}
\usepackage{esint}
\usepackage{siunitx}
\usepackage[unicode=true,pdfusetitle,
 bookmarks=true,bookmarksnumbered=false,bookmarksopen=false,
 breaklinks=false,pdfborder={0 0 1},backref=false,colorlinks=false]
 {hyperref}
\usepackage{breakurl}

\makeatletter

\@ifundefined{textcolor}{}
{%
 \definecolor{BLACK}{gray}{0}
 \definecolor{WHITE}{gray}{1}
 \definecolor{RED}{rgb}{1,0,0}
 \definecolor{GREEN}{rgb}{0,1,0}
 \definecolor{BLUE}{rgb}{0,0,1}
 \definecolor{CYAN}{cmyk}{1,0,0,0}
 \definecolor{MAGENTA}{cmyk}{0,1,0,0}
 \definecolor{YELLOW}{cmyk}{0,0,1,0}
}

\usepackage{babel}

\makeatother

\begin{document}

\title{Proximity effect in graphene-topological insulator heterostructures}

\author{Junhua Zhang, C. Triola, and E. Rossi}

\affiliation{Department of Physics, College of William and Mary, Williamsburg,
Virginia 23187, USA}

\date{\today}
\begin{abstract}
We formulate a continuum model to study the low-energy electronic
structure of heterostructures formed by graphene on a strong three-dimensional
topological insulator (TI) for the case of both commensurate and incommensurate
stacking. The incommensurability can be due to a twist angle between
graphene and the TI surface or a lattice mismatch between the two systems.  
We find that the proximity of the TI induces in graphene a strong enhancement of the 
spin-orbit coupling that can be tuned via the twist angle.

\end{abstract}

\maketitle

The surfaces of strong three-dimensional (3D) topological insulators
(TIs) \cite{hasan2010} and graphene \cite{novoselov2004,neto2009}
have very similar low-energy electronic structures:
the conduction and valence bands touch at isolated points, the
Dirac points (DPs), and around these points the fermionic excitations
are well described as massless two-dimensional (2D) chiral Dirac
fermions for which the phase of a two-state quantum degree of freedom
is locked with the momentum direction. However, there are also qualitative
differences: (i) in graphene the chirality is associated with the
sublattice degree of freedom whereas in a TI surface (TIS) it is associated
with the electron spin; (ii) in graphene the number of DPs is even
whereas in a TIS it is odd ; (iii) in TIs the electron-phonon scattering
is much stronger than that in graphene. Therefore, the
transport properties of graphene \cite{dassarma2011} and TIs are
different in significant aspects: in graphene, because the
intrinsic spin-orbit (SO) coupling is negligible \cite{min2006,yao2007,gmitra2009,rashba2009},
no quantum spin Hall effect is expected, contrary to the case in a TI; graphene
has the highest room-temperature mobility, whereas
TIs have very low mobilities. 
These facts, together with the recent experimental progress \cite{dang2010, *song2010},
motivated us to study graphene-TI heterostructures, in which
the proximity to a TI is expected to enhance the SO coupling of graphene
and create a novel 2D system with nontrivial spin textures and high,
room-temperature, electron mobility. This approach to enhance the SO
coupling in graphene appears to be more practical than previously
proposed approaches \cite{weeks2011, *jiang2012}
that rely on doping graphene with heavy adatoms.

In this work, we study the low-energy electronic structure of heterostructures
formed by one sheet of graphene placed on the conducting surface of
a 3D TI. 
Our results show that not only can the proximity of a TIS
enhance the SO coupling in graphene and bilayer graphene (BLG), but also
that this enhancement can be controlled via the relative rotation,
the twist angle, between the graphene lattice and the TI's lattice.
The presence of a relative rotation, in general, induces an incommensurate
stacking of the graphene and the substrate 
\cite{novoselov2005b, *kim2009, *kim2010, *ponomarenko2011, *kim2012, *gorbachev2012, *haigh2012, *jzhang2013}
\cite{dossantos2007, mele2010, shallcross2010, suarez2010, bistritzer2010, li2010, mele2011, bistritzer2011b, bistritzer2011, morell2011, mele2012, kindermann2011,luican2011,yan2012,sanchez2012,sanjose2013,kumar2013,lu2014,dean2010,xue2011,yankowitz2012,ponomarenko2013,hunt2013,dean2013,mucha2013}. 
As a consequence  
we develop and present a theory that is able to take 
into account the incommensurability between graphene and the TIS.
This cannot be achieved via standard approaches, such as density functional
theory \cite{jin2013}, and tight-binding models, due to the computational cost
of these approaches for incommensurate structures. 
A continuous model, on the other hand, can effectively
treat heterostructures with incommensurate stacking. To develop the theory for incommensurate structures, 
however, we need a continuous model for the commensurate limit.
We present such a model and then,
starting from it, the model able to treat incommensurate graphene-TI structures.
Our results show that in graphene-TI heterostructures the proximity effect
induces a strong enhancement of the SO coupling in graphene, nontrivial
spin and pseudospin textures on the bands, and that all these effects can
be tuned to great extent via the relative rotation between graphene and the TI.
Moreover, we present results for the case in which  tunneling processes with finite momentum transfer are present.

We consider the TI material to be a tetradymite
such as $\mathrm{Bi_{2}Se_{3}}$, $\mathrm{Bi_{2}Te_{3}}$, and $\mathrm{Sb_{2}Te_{3}}$.
In these compounds the surface states are
found on the 111 surface. 
The projected surface Brillouin zone (BZ) is hexagonal with
a single DP at the zone center \cite{zhang2009ti}. Let
$a_{2}$ be the effective lattice constant that corresponds to the
surface BZ and 
$a_1 = \SI{2.46}{\angstrom}$ the graphene lattice constant.
We have $a_{2}/(\sqrt{3}a_{1})=1+\delta$ with $\delta<1\%$
for $\mathrm{Sb_{2}Te_{3}}$ and $\delta\approx-3\%$ ($\delta\approx+3\%$
) for $\mathrm{Bi_{2}Se_{3}}$ ($\mathrm{Bi_{2}Te_{3}}$). Thus,
the study of the commensurate $\sqrt{3}\times\sqrt{3}$ stacking pattern
is expected to be a good approximation for graphene-$\mathrm{Sb_{2}Te_{3}}$ heterostructure
and for developing the theory for incommensurate structures.  
The Hamiltonian describing the electronic degrees of freedom of the
heterostructure can be written as $H=H^{\text{g}}+H^{\text{TIS}}+H_{t}$,
where $H^{\text{g}}$ is the Hamiltonian for an isolated sheet of
graphene, $H^{\text{TIS}}$ is the Hamiltonian for the TIS, and $H_{t}$
describes tunneling processes between graphene
and the TIS. The long wavelength physics of graphene is described
by a pair of 2D massless Dirac Hamiltonians: $H^{\text{g },K}=\sum_{\mathbf{p},\sigma,\tau\tau'}c_{\mathbf{K}+\mathbf{p},\tau,\sigma}^{\dagger}\left(\hbar v_{1}\boldsymbol{\tau}\cdot\mathbf{p}-\mu_{1}\right)_{\tau\tau'}c_{\mathbf{K}+\mathbf{p},\tau',\sigma}$
and $H^{\text{g},K'}=\sum_{\mathbf{p},\sigma,\tau\tau'}c_{\mathbf{K}'+\mathbf{p},\tau,\sigma}^{\dagger}\left(\hbar v_{1}\boldsymbol{\tau}^{*}\cdot\mathbf{p}-\mu_{1}\right)_{\tau\tau'}c_{\mathbf{K}'+\mathbf{p},\tau',\sigma}$,
where $c_{\mathbf{K}+\mathbf{p},\tau,\sigma}^{\dagger}$ ($c_{\mathbf{K}+\mathbf{p},\tau,\sigma}$)
creates (annihilates) a Dirac fermion on sublattice $\tau\,(A,B)$
with spin $\sigma\,(\uparrow,\downarrow)$ at a Dirac wave vector
$\mathbf{p}$ measured from one of the two inequivalent BZ corners
($K$ and $K'$ valley) located at wave vectors $\mathbf{K}$ and
$\mathbf{K}'$ ($|\mathbf{p}|\ll|\mathbf{K}|$), $\boldsymbol{\tau}=\left(\begin{array}{cc}
\tau^{x}, & \tau^{y}\end{array}\right)$ are Pauli matrices acting on the sublattice space, $v_{1}\approx10^{6}\mathrm{m/s}$
is the Fermi velocity, and $\mu_{1}$ is the chemical potential.
The TIS states near its Dirac point can be described by an effective 2D continuum
model \cite{zhang2009ti,liucx2010}: $H^{\text{TIS}}=\sum_{\mathbf{k},\sigma\sigma'}a_{\mathbf{k},\sigma}^{\dagger}\left[\hbar v_{2}\left(\boldsymbol{\sigma}\times\mathbf{k}\right)\cdot\hat{\mathbf{z}}-\mu_{2}\right]_{\sigma\sigma'}a_{\mathbf{k},\sigma'}$,
where $a_{\mathbf{k},\sigma}^{\dagger}$ ($a_{\mathbf{k},\sigma}$)
creates (annihilates) a surface massless Dirac fermion with spin $\sigma$
at wave vector $\mathbf{k}$ measured from the zone center ($\overline{\Gamma}$-point),
$\boldsymbol{\sigma}=\left(\begin{array}{cc}
\sigma^{x}, & \sigma^{y}\end{array}\right)$ are Pauli matrices acting on spin space, $\hat{\mathbf{z}}$ is the
unit vector along the $z$ direction, and $\mu_{2}$ is the chemical potential.
In $\mathrm{Bi_{2}Se_{3}}$,
$\mathrm{Bi_{2}Te_{3}}$, and $\mathrm{Sb_{2}Te_{3}}$, the Fermi
velocity $v_{2}$ is roughly half of that in graphene; hence, in the
remainder we assume $v_{2}=v_{1}/2$. 
In our model we neglect the hexagonal warping of the TIS bands
due to higher-order terms in $k$ in $H^{\rm TIS}$ \cite{fu2009b}. The reason is that such
effects are non-negligible only at relative high energies $\gtrsim 200$~meV
away from the TI's DP \cite{chen2009,fu2009b} and we are only interested
in the energy range close to the TI's DP. 
We also neglect effects due to the TI's bulk states \cite{analytis2010} for two reasons:
(i) 
in current experiments the effect of the bulk states can be strongly suppressed
via chemical and field effect doping \cite{chen2009,hsieh2009,steinberg2010,checkelsky2011}, 
and by using TI's thin films \cite{zhangy2010,taskin2012};
(ii) 
the most interesting situation arises when the bulk states can be neglected: in this
case the properties of the systems are dominated not by the TI's bulk states but by the states resulting from the hybridization
of the graphene and the TI's surface states.
The form of
$H_{t}$ depends on the stacking pattern and the interface properties as we show below.

\begin{figure}[tp]
\includegraphics[width=8.5cm]{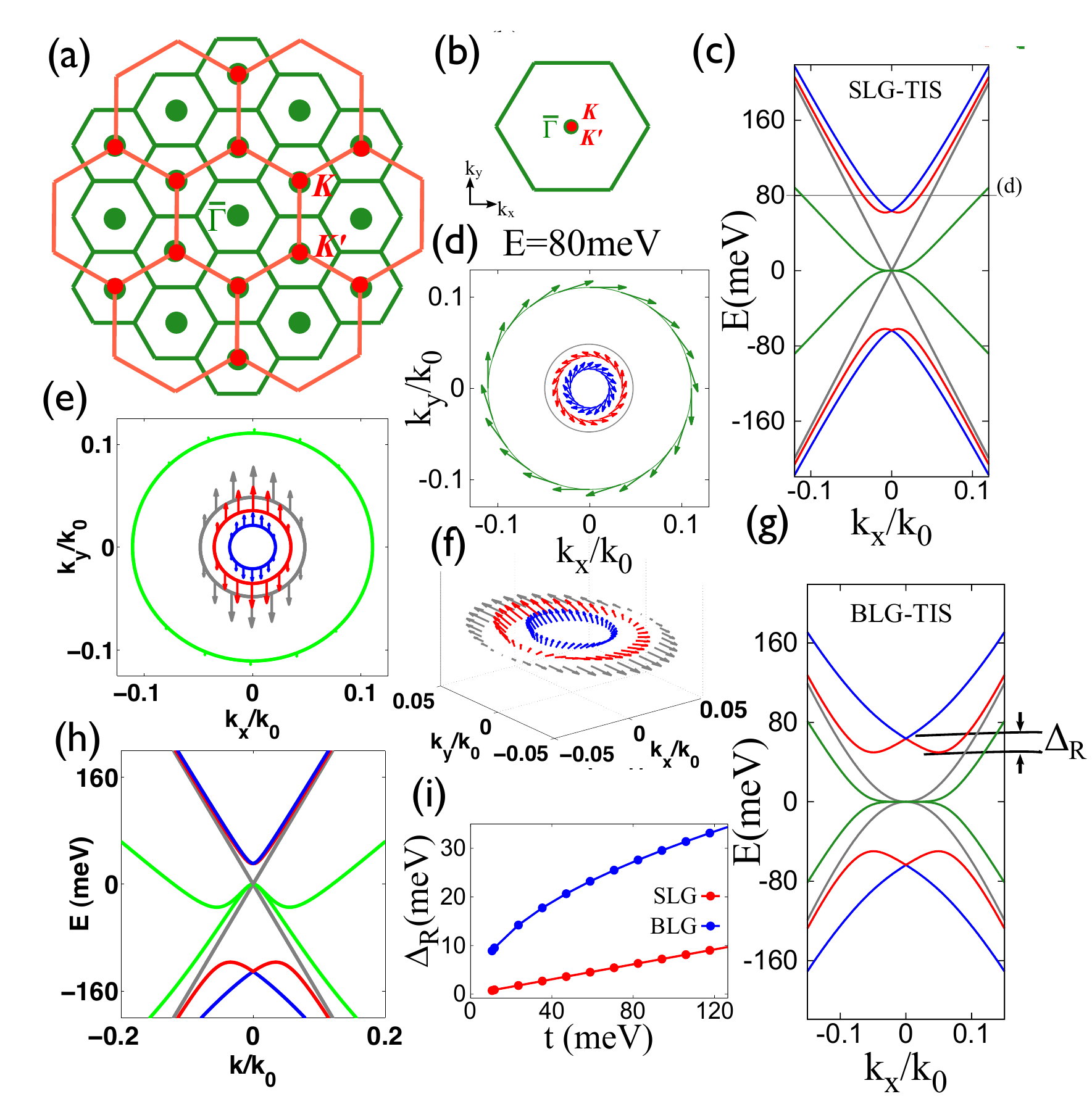}
\caption{(Color online) 
(a) Schematics of the $\sqrt{3}\times\sqrt{3}$ stacked
graphene BZ (red or dark) and TIS BZ (green or light) in the repeated zone scheme without
tunneling. 
(b) Folded BZ after turning on tunneling. 
(c) Renormalized
bands of SLG-TIS for $\mu_1=\mu_2=0$. Here $k_0\equiv 830\text{meV}/(\hbar v_2)$. 
(d) Spin texture on
the bands at $E=80\,\text{meV}$. The arrows indicate spin directions. 
(e) Texture of the in-plane component of the pseudospin at $E=80\,\text{meV}$, 
(f) shows the full pseudospin orientation on the three Fermi surfaces closest to the $\bar\Gamma$ point.
(g) Renormalized bands of BLG-TIS. 
(h) Renormalized
bands of SLG-TIS for $\mu_1=0$, $\mu_2=100$~meV.
(i) Rashba-like splitting $\Delta_{\text{R}}$ in SLG-TIS and BLG-TIS
as a function of $t$. \label{fig1}}.
\end{figure}

We first consider the graphene-TI heterostructure in a $\sqrt{3}\times\sqrt{3}$
commensurate stacking, in which each TIS atom is directly
underneath a carbon atom. The strongest tunneling is expected to occur
between the directly stacked atoms, among which all the carbon atoms
can be shown to belong to one sublattice (e.g., sublattice A). 
As a result of the periodic tunneling potential,
in the BZ of the heterostructure the original graphene BZ is folded such
that the two valleys are both located at the zone center overlapping
with the TIS DP, Fig.\,\ref{fig1}~(a),~(b). In this case the tunneling Hamiltonian can be written as
$H_{t}=\sum_{\mathbf{k},\lambda,\tau,\sigma}t_{\tau}a_{\mathbf{k},\sigma}^{\dagger}c_{\lambda,\mathbf{k},\tau,\sigma}+h.c.$,
where $\lambda=K,K'$ and the tunneling matrix elements $t_{A}=t,\ t_{B}=0$
are assumed to be spin and momentum independent. The Hamiltonian for
such a structure takes the form
\begin{equation}
\hat{\mathcal{H}}_{\mathbf{k}}=\left(\begin{array}{ccc}
\hat{H}_{\mathbf{k}}^{\text{g},K} & 0 & \hat{T}^{\dagger}\\
0 & \hat{H}_{\mathbf{k}}^{\text{g},K'} & \hat{T}^{\dagger}\\
\hat{T} & \hat{T} & \hat{H}_{\mathbf{k}}^{\text{TIS}}
\end{array}\right),\ \ \hat{T}=\left(\begin{array}{cccc}
t & 0 & 0 & 0\\
0 & 0 & t & 0
\end{array}\right),\label{eq: commensurate_Hamiltonian}
\end{equation}
where the graphene blocks are $4\times4$ matrices in sublattice and
spin space whereas the TIS block is a $2\times 2$ matrix in spin space.

Insights can be achieved using a perturbative approach \cite{hutasoit2011}.
In this approach the effect of tunneling processes on the graphene spectrum is captured
by the self-energy  $\hat{\Sigma}_{\mathbf{k}}(i\omega_{n})=\hat{V}^{\dagger}\hat{G}_{\mathbf{k}}^{0}(i\omega_{n})\hat{V}$,
where $\hat{G}_{\mathbf{k}}^{0}(i\omega_{n})$
is the Green's function of the TIS and $\hat{V}$ is the tunneling vertex. 
In the basis formed by
the eigenstates of the Hamiltonian of isolated graphene, $\Phi_{\lambda,\mathbf{k},\alpha,\sigma}$,
where $\alpha=\pm$ refer to the fourfold degenerate upper and lower bands  we obtain
\begin{align}
\hat{\Sigma}_{\mathbf{k}}(i\omega_{n})= & \left(\begin{array}{cc}
\Sigma_{\mathbf{k}}^{S}(i\omega_{n}) & e^{-i\left(\theta_{\mathbf{k}}-\frac{\pi}{2}\right)}\Sigma_{\mathbf{k}}^{A}(i\omega_{n})\\
e^{i\left(\theta_{\mathbf{k}}-\frac{\pi}{2}\right)}\Sigma_{\mathbf{k}}^{A}(i\omega_{n}) & \Sigma_{\mathbf{k}}^{S}(i\omega_{n})
\end{array}\right)\nonumber \\
 & \otimes\left(\mathrm{I}_{\alpha}+\sigma_{\alpha}^{x}\right)\otimes\left(\mathrm{I}_{\lambda}+\sigma_{\lambda}^{x}\right),\label{eq: comm_self_energy}
\end{align}
where $\Sigma_{\mathbf{k}}^{S/A}(i\omega_{n})=\left(t^{2}/2\right)G_{\mathbf{k}}^{S/A}(i\omega_{n})$
with $G_{\mathbf{k}}^{S/A}(i\omega_{n})=\left[1/\left(i\omega_{n}-\hbar v_{2}k+\mu_{2}\right)\pm1/\left(i\omega_{n}+\hbar v_{2}k+\mu_{2}\right)\right]/2$,
and the first $2\times2$ matrix
acts in the spin space, $\left(\mathrm{I}_{\alpha}+\sigma_{\alpha}^{x}\right)$
acts in the band space, and $\left(\mathrm{I}_{\lambda}+\sigma_{\lambda}^{x}\right)$
in the valley space. $\mathrm{I}$ is the  $2\times2$ identity
matrix and $\theta_{\mathbf{k}}=\arctan(k_{y}/k_{x})$.
The appearance of nonzero off-diagonal spin components
with phase factor $\left(\theta_{\mathbf{k}}-\frac{\pi}{2}\right)$
in the self-energy indicates an induced helical spin texture on some of
the graphene bands. The renormalized graphene bands in the perturbative
approach coincide with those obtained by direct diagonalization.
Figure \ref{fig1}(c) shows
the band structure of a graphene-TI heterostructure with $t=45\,\text{meV}$ and $\mu_{1}=\mu_{2}=0$.
We see that the fourfold degeneracy of the original graphene bands is
partially lifted.
In our model the eigenstates of the hybridized bands can be calculated explicitly.
This allows us to:
(i)
obtain directly both the spin and {\em pseudospin} configuration on all the
renormalized bands, Fig.~\ref{fig1}~(d)-(f); 
(ii)
show that,
as expected from the form of the self-energy,
on the two gapped bands (forming the two smaller Fermi surfaces)
the in-plane spin is locked perpendicular to the momentum and
winds around the $\overline{\Gamma}$ point either clockwise or counterclockwise,
analogous to a system with Rashba-type SO coupling, Fig.~\ref{fig1}~(d); 
(iii) show that the spin helicity of the hybridized bands can be different (opposite) to the helicity of the original TI's band, Fig.~\ref{fig1}~(d);
(iv)
show that the two degenerate bands, seemingly equal to the original graphene (or BLG) bands, 
are in reality  antisymmetric combinations of the states of isolated graphene (or BLG)
at opposite valleys: $\frac{1}{\sqrt{2}}\left(\Phi_{K,\alpha,\uparrow}-\Phi_{K',\alpha,\uparrow}\right)$
and $\frac{1}{\sqrt{2}}\left(\Phi_{K,\alpha,\downarrow}-\Phi_{K',\alpha,\downarrow}\right)$,
$\alpha=\pm$;
and (v) 
show that the two degenerate bands have a unique pseudospin structure, very different from the pseudospin structure of both
the original $K$ and $K'$ valleys, which we expect would affect transport measurements, Fig.~\ref{fig1}~(e),~(f).
In addition, our model is easily generalized to the case of BLG. 
The results of Fig.~\ref{fig1}~(g) show the bands of a BLG-TI heterostructure
and reveal that the enhancement, due to the proximity effect, of the SO coupling in BLG
is much larger than in single layer graphene (SLG), Fig.~\ref{fig1}~(i). This is due mostly
to the fact that, at low energies, BLG has a much higher density of states (DOS) than SLG.
Finally, we consider the effect of
a difference $\delta\mu=\mu_2-\mu_1$ between the TI's 
and  graphene chemical potential. 
By varying $\delta\mu$ the value of $\bf k$ for which the pristine bands of the TI and graphene cross,
and for which the hybridization is stronger, can be  tuned. 
Figure \ref{fig1}~(h)
shows the case for which $\mu_2=100$~meV and $\mu_1=0$. We see that in this
case the induced Rashba splitting is stronger than when $\mu_2=\mu_2=0$.
This is due to the fact that the DOS increases as we move away from the DP.

\begin{figure}[tp]
\includegraphics[width=8.6cm]{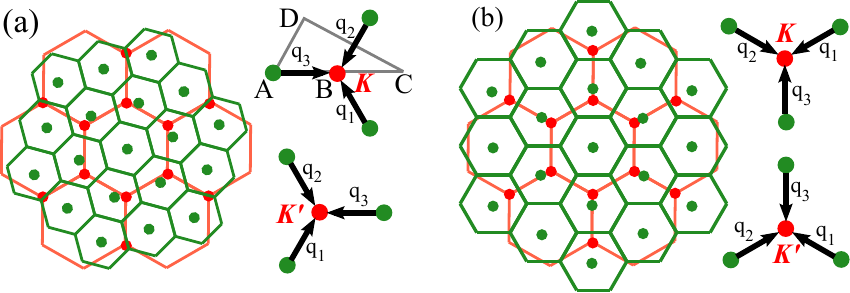}
\caption{(Color online) Schematics of the graphene and TIS BZs in an incommensurate
structure formed from (a) a twist (b) a lattice mismatch,
with the corresponding $\mathbf{q}_{j}$ vectors at the $K$ and
$K'$ points. \label{fig2}}
\end{figure}

We now consider incommensurate structures.
The tunneling matrix elements can be written as: 
\begin{equation}
T_{\tau}(\mathbf{k}_{2},\mathbf{k}_{1})=\sum_{\mathbf{G}_{1},\mathbf{G}_{2}}\frac{t(\mathbf{k}_{1}+\mathbf{G}_{1})}{\sqrt{3}\Omega_{1}}e^{i\mathbf{G}_{1}\cdot\mathbf{d}_{\tau}}\delta_{\mathbf{k}_{2}+\mathbf{G}_{2},\mathbf{k}_{1}+\mathbf{G}_{1}}\label{eq: tunneling_general}
\end{equation}
where the crystal momentum is conserved by the tunneling process in
which a graphene quasiparticle of wave vector $\mathbf{k}_{1}$
residing on sublattice $\tau$ hops to a TIS state with wave vector
$\mathbf{k}_{2}$.  
$\Omega_{1}$ is the graphene unit
cell area and $\mathbf{d}_{A}=\mathbf{0}$, $\mathbf{d}_{B}=\left(\begin{array}{cc}
-a_{0}, & 0\end{array}\right)$ are the positions of the two carbon atoms in a unit cell with carbon-carbon
distance $a_{0}$. $\{\mathbf{G}_{1}\},\{\mathbf{G}_{2}\}$ are the
reciprocal lattice vectors of graphene and TIS, respectively.
$t(\mathbf{k})$ are the Fourier amplitudes of the tunneling potential
$t(\mathbf{r})$ assumed to be a smooth function of $\mathbf{r}$,
the spatial separation between graphene and TIS atoms projected onto
the interface plane.
Given that the graphene-TIS separation distance exceeds the interatomic
distance in each material, the dominant tunneling amplitudes of $t(\mathbf{k})$
near the graphene DP are the ones with $|\mathbf{k}|=K_{D}\equiv\left|\mathbf{K}\right|$.
This allows to restrict the
sum over $\{\mathbf{G}_{1}\}$ to three vectors: $\mathbf{g}_{1}(=\mathbf{0}),\ \mathbf{g}_{2},\ \mathbf{g}_{3}$,
where the latter two connect a valley with its equivalent first BZ
corners. For small wave vectors measured from the respective DPs,
we have 
$H_{t}=\sum_{\mathbf{p},\tau,\sigma}\sum_{j,l,\dots=1}^{3}[T_{\tau,j}a_{\mathbf{p}+\mathbf{q}_{j},\sigma}^{\dagger}c_{\mathbf{p},\tau,\sigma}+T_{\tau,l}^{*}c_{\mathbf{p}+\mathbf{q}_{j}+\bar{\mathbf{q}}_{l},\tau,\sigma}^{\dagger}a_{\mathbf{p}+\mathbf{q}_{j},\sigma}+\dots]$,
where $T_{\tau,j}=t'e^{i\mathbf{g}_{j}\cdot\mathbf{d}_{\tau}}$ with
$t'\equiv t(K_{D})/\left(\sqrt{3}\Omega_{1}\right)$, $\{\mathbf{q}_{j}\}$
are the offset vectors between the graphene DP and the three ``nearest-neighboring''
TIS DPs, and $\bar{\mathbf{q}}_{l}\in\{-\mathbf{q}_{j}\}$, as shown in Fig.\,\ref{fig2}. 
The repeated action of the ``nonlocal'' coupling generates
a $k$-space lattice \cite{bistritzer2011}.  
For a rotation angle $\theta$, the separation between the offset
DPs is $|\mathbf{q}_{j}|\equiv q=2K_{D}\sin(\theta/2)$,
for the lattice mismatch case, 
$q=\left|\delta/(1+\delta)\right|K_{D}$, Fig.~\ref{fig2}.

For very small twist angles or lattice mismatches such that the dimensionless
parameter $\gamma\equiv\frac{t'}{\hbar v_{2}q}>1$, graphene and TIS
will be strongly coupled. However, when $\gamma<1$, a weak coupling
theory is valid \cite{dossantos2007,bistritzer2011,mele2012}. In
this case, to investigate the low-energy spectrum of graphene, we
can truncate the $k$-space lattice and obtain the Hamiltonian: 
\begin{align}
\hat{\mathcal{H}}_{\mathbf{p}} & =\left(\begin{array}{cccc}
\hat{H}_{\mathbf{p}}^{\text{g},K} & \hat{T}_{1}^{\dagger} & \hat{T}_{2}^{\dagger} & \hat{T}_{3}^{\dagger}\\
\hat{T}_{1} & \hat{H}_{\mathbf{q}_{1}+\mathbf{p}}^{\text{TIS}} & 0 & 0\\
\hat{T}_{2} & 0 & \hat{H}_{\mathbf{q}_{2}+\mathbf{p}}^{\text{TIS}} & 0\\
\hat{T}_{3} & 0 & 0 & \hat{H}_{\mathbf{q}_{3}+\mathbf{p}}^{\text{TIS}}
\end{array}\right),\label{eq: incomm_Hamiltonian}\\
\hat{T}_{1} & =\left(\begin{array}{cccc}
t' & t' & 0 & 0\\
0 & 0 & t' & t'
\end{array}\right),\ \hat{T}_{2}=\left(\begin{array}{cccc}
t' & t'e^{-i\frac{2\pi}{3}} & 0 & 0\\
0 & 0 & t' & t'e^{-i\frac{2\pi}{3}}
\end{array}\right),\nonumber \\
\hat{T}_{3} & =\left(\begin{array}{cccc}
t' & t'e^{i\frac{2\pi}{3}} & 0 & 0\\
0 & 0 & t' & t'e^{i\frac{2\pi}{3}}
\end{array}\right).\nonumber 
\end{align}
A similar Hamiltonian is valid for the $K'$ valley \cite{note2013}.
In the absence of twist and mismatch, the system reduces to the commensurate
structure, giving rise to
$\hat{T}=\hat{T}_{1}+\hat{T}_{2}+\hat{T}_{3}$, so that $t'=t/3$.

\begin{figure}[tp]
\includegraphics[width=8.5cm]{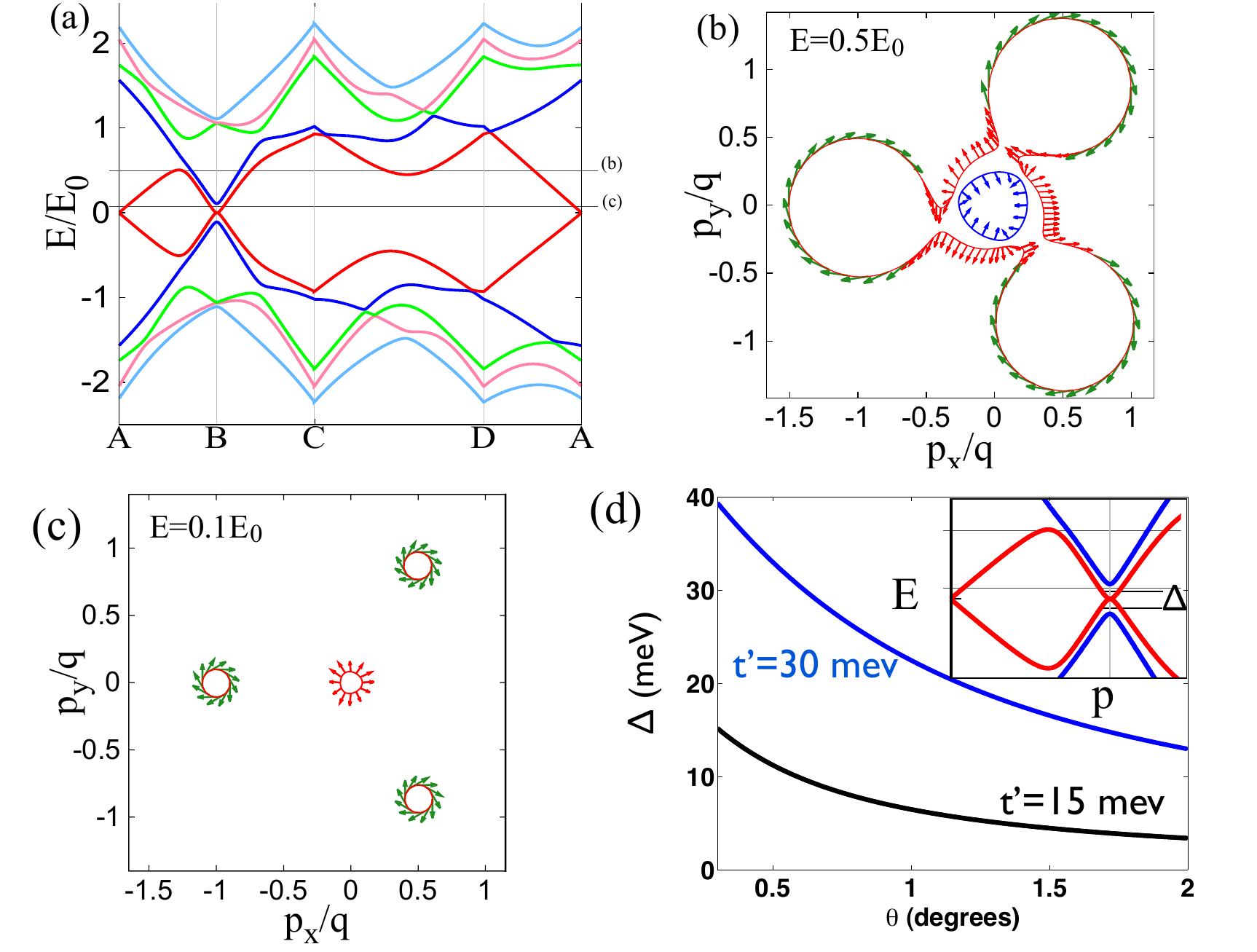}
\caption{(Color online) (a) The band structure along the path A-B-C-D-A indicated
in Fig.\,\ref{fig2}(a).
(b),(c) show the spin texture on the bands at different energies.
$E_{0}\equiv\hbar v_{2}q=t'/\gamma$.
(d) Splitting ($\Delta)$ of the low-energy bands as a function of twist angle
    for $t'=30$~meV  and $t'=15$~meV. 
\label{fig3}}
\end{figure}

Figures \ref{fig3}~(a)-(c) show the band and spin structure around the $K$ point 
for  an incommensurate graphene-TI heterostructure with $\gamma=0.2$, $t'=15\,\text{meV}$ and $\mu_{1}=\mu_{2}=0$.
The result for the $K'$ point is simply a $60^\circ$ rotation
of the former. 
The results of Fig.~\ref{fig3} show that: 
(i) 
the original twofold spin degeneracy of the graphene Dirac cone is completely lifted; 
(ii)
of the two original degenerate linear bands one is now fully gapped and the other is no
longer linear at the DP; 
(iii) 
the bands acquire nontrivial in-plane spin textures. 
The key property of graphene-TI heterostructures 
is that the features of the band structure and spin texture can
be controlled via the twist angle. By changing the value of $\theta$, for fixed
$t'$ and energy, the distance between the Fermi pockets shown in Fig.~\ref{fig3}~(b) and (c),
and their size, can be tuned. In addition, the splitting of the low-energy bands $\Delta$
can be controlled as shown in Fig.~\ref{fig3}~(d).

\begin{figure}[tp]
\includegraphics[width=8.5cm]{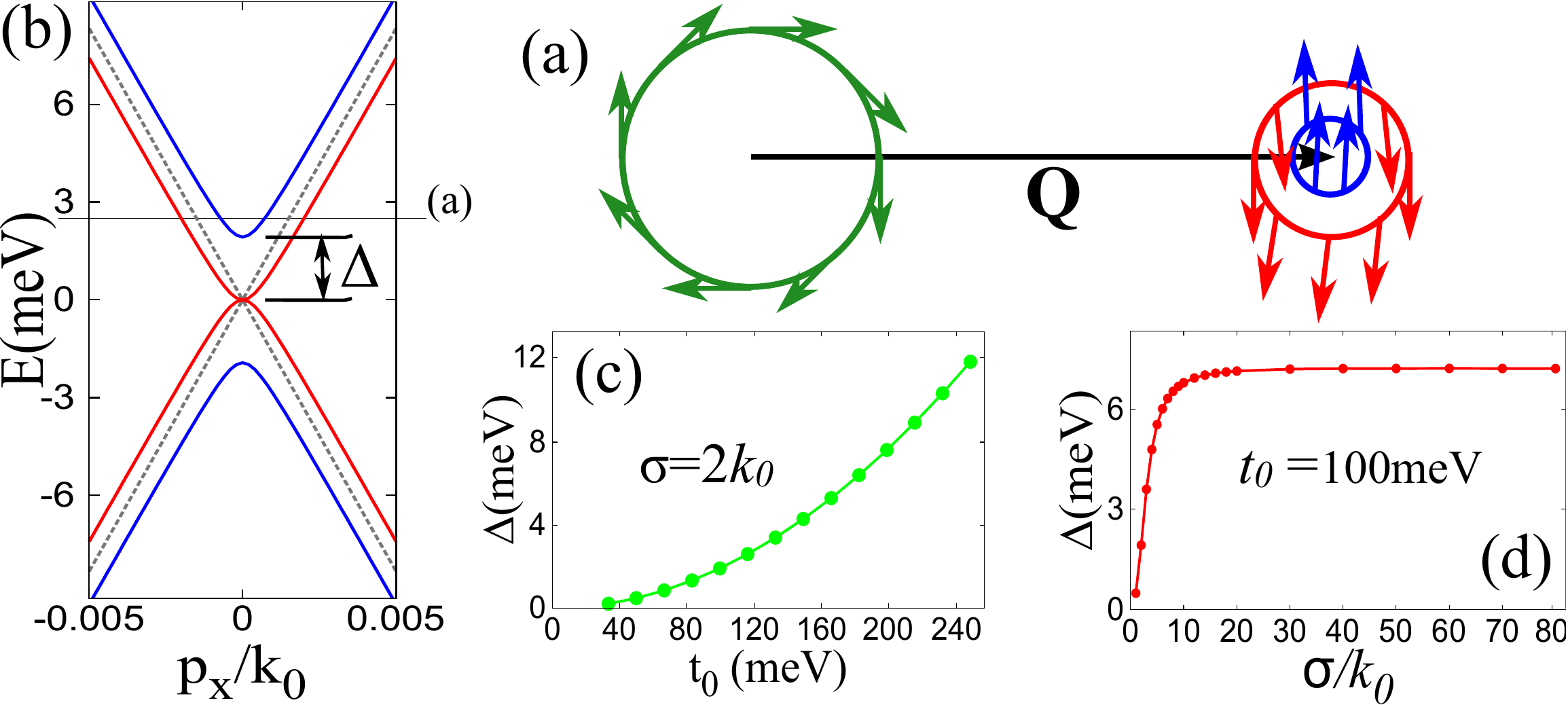}
\caption{(Color online)  (a) Schematics of the induced spin texture on graphene
(right) from the TIS spin helix (left).
(b) Renormalized graphene bands (solid lines)
for $t_{0}=100\,\text{meV}$, $\sigma=2k_0$.
Spin-split gap ($\Delta$) as a function of (c) $t_{0}$ and (d) $\sigma$.
\label{fig4}}
\end{figure}

In the presence of surface roughness and/or phonons 
tunneling processes with finite momentum transfer are allowed. We expect
the effect of such processes to be weak, however,
to gain some insight, we consider the case in which
the tunneling amplitude has a Gaussian profile with respect to the
momentum transfer $\mathbf{q}$: $t_{\mathbf{q}}=t_{0}\exp\left(-|\mathbf{q}|^{2}/(2\sigma^{2})\right)$,
where $t_{0}$ characterizes the tunneling strength and $\sigma$
the variance. To qualitatively understand the effect of such processes,
we study the case of an isolated graphene Dirac cone separated by
a large wave vector $\mathbf{Q}$ from the closest TIS DP.
With the use of the perturbative approach outlined above, the proximity effect
on the graphene spectrum is captured by the self-energy 
\begin{align}
 & \hat{\Sigma}_{\mathbf{Q}+\mathbf{p}}(i\omega_{n})=\left(\mathrm{I}_{\alpha}+\sigma_{\alpha}^{x}\right)\otimes\nonumber \\
 & \left(\begin{array}{cc}
\Sigma_{\mathbf{Q}+\mathbf{p}}^{S}(i\omega_{n}) & e^{-i\left(\theta_{\mathbf{Q}+\mathbf{p}}-\frac{\pi}{2}\right)}\Sigma_{\mathbf{Q}+\mathbf{p}}^{A}(i\omega_{n})\\
e^{i\left(\theta_{\mathbf{Q}+\mathbf{p}}-\frac{\pi}{2}\right)}\Sigma_{\mathbf{Q}+\mathbf{p}}^{A}(i\omega_{n}) & \Sigma_{\mathbf{Q}+\mathbf{p}}^{S}(i\omega_{n})
\end{array}\right)\label{eq: average_self_energy}
\end{align}
with $\Sigma_{\mathbf{Q}+\mathbf{p}}^{S/A}(i\omega_{n})=\frac{t_{0}^{2}\Omega_{2}}{2\pi}\exp\left[-\frac{|\mathbf{Q+p}|^{2}}{\sigma^{2}}\right]\int_{0}^{\infty}k\exp\left[-\frac{k^{2}}{\sigma^{2}}\right]\times$
$I_{0/1}(\frac{2|\mathbf{Q+p}|}{\sigma^{2}}k)G_{\mathbf{k}}^{S/A}(i\omega_{n})\mathrm{d}k$,
where $I_{n}(x),\ n=0,1$ are the modified Bessel functions of the
first kind. The form of the phase factors in the off-diagonal spin
components of $\hat{\Sigma}$ implies an induced spin texture on graphene
with the spin perpendicular to the wave vector $\mathbf{Q+p}$,
Fig.\ref{fig4}~(a). We find, Fig.\,\ref{fig4}(b), that also in
this case the spin degenerate bands are split and the remaining
gapless bands are no longer linear. Figures \ref{fig4}(c) and (d)
show the size of the gap between spin-split bands as a function of
$t_{0}$ and $\sigma$, respectively.

In conclusion, we have studied the proximity effect of a strong 3D
TI on the low-energy spectrum of graphene in commensurate and
incommensurate structures as well as in a case with surface roughness.
To be able to take into account the incommensurability we have
developed a continuous model. Using this model we have been able to identify, 
both for commensurate and incommensurate stacking, the spin and pseudospin structure
of all the hybridized bands and show that it is very
unusual and likely to affect transport measurements. We have also
shown that the enhancement of the SO coupling is in general much stronger
in BLG than graphene. 
In addition, we have shown that properties of these bands, and their
spin structures, can be substantially tuned by varying 
the relative rotation between the graphene lattice and the TI's lattice.

We acknowledge helpful discussions with Michael Fuhrer and Elsa Prada.
This work is supported by ONR, Grant No. ONR-N00014-13-1-0321,
ACS-PRF No. 53581-DNI5, and
the Jeffress Memorial Trust. C.T. acknowledges support
from the Virginia Space Grant Consortium.


\begin{thebibliography}{58}
\expandafter\ifx\csname natexlab\endcsname\relax\def\natexlab#1{#1}\fi
\expandafter\ifx\csname bibnamefont\endcsname\relax
  \def\bibnamefont#1{#1}\fi
\expandafter\ifx\csname bibfnamefont\endcsname\relax
  \def\bibfnamefont#1{#1}\fi
\expandafter\ifx\csname citenamefont\endcsname\relax
  \def\citenamefont#1{#1}\fi
\expandafter\ifx\csname url\endcsname\relax
  \def\url#1{\texttt{#1}}\fi
\expandafter\ifx\csname urlprefix\endcsname\relax\def\urlprefix{URL }\fi
\providecommand{\bibinfo}[2]{#2}
\providecommand{\eprint}[2][]{\url{#2}}

\bibitem[{\citenamefont{Hasan and Kane}(2010)}]{hasan2010}
\bibinfo{author}{\bibfnamefont{M.~Z.} \bibnamefont{Hasan}} \bibnamefont{and}
  \bibinfo{author}{\bibfnamefont{C.~L.} \bibnamefont{Kane}},
  \bibinfo{journal}{Rev. Mod. Phys.} \textbf{\bibinfo{volume}{82}},
  \bibinfo{pages}{3045} (\bibinfo{year}{2010}).

\bibitem[{\citenamefont{Novoselov et~al.}(2004)\citenamefont{Novoselov, Geim,
  Morozov, Jiang, Zhang, Dubonos, Grigorieva, and Firsov}}]{novoselov2004}
\bibinfo{author}{\bibfnamefont{K.~S.} \bibnamefont{Novoselov}},
  \bibinfo{author}{\bibfnamefont{A.~K.} \bibnamefont{Geim}},
  \bibinfo{author}{\bibfnamefont{S.~V.} \bibnamefont{Morozov}},
  \bibinfo{author}{\bibfnamefont{D.}~\bibnamefont{Jiang}},
  \bibinfo{author}{\bibfnamefont{Y.}~\bibnamefont{Zhang}},
  \bibinfo{author}{\bibfnamefont{S.~V.} \bibnamefont{Dubonos}},
  \bibinfo{author}{\bibfnamefont{I.~V.} \bibnamefont{Grigorieva}},
  \bibnamefont{and} \bibinfo{author}{\bibfnamefont{A.~A.}
  \bibnamefont{Firsov}}, \bibinfo{journal}{Science}
  \textbf{\bibinfo{volume}{306}}, \bibinfo{pages}{666} (\bibinfo{year}{2004}).

\bibitem[{\citenamefont{Neto et~al.}(2009)\citenamefont{Neto, Guinea, Peres,
  Novoselov, and Geim}}]{neto2009}
\bibinfo{author}{\bibfnamefont{A.~H.~C.} \bibnamefont{Neto}},
  \bibinfo{author}{\bibfnamefont{F.}~\bibnamefont{Guinea}},
  \bibinfo{author}{\bibfnamefont{N.~M.~R.} \bibnamefont{Peres}},
  \bibinfo{author}{\bibfnamefont{K.~S.} \bibnamefont{Novoselov}},
  \bibnamefont{and} \bibinfo{author}{\bibfnamefont{A.~K.} \bibnamefont{Geim}},
  \bibinfo{journal}{Rev. Mod. Phys.} \textbf{\bibinfo{volume}{81}},
  \bibinfo{pages}{109} (\bibinfo{year}{2009}).

\bibitem[{\citenamefont{{Das Sarma} et~al.}(2011)\citenamefont{{Das Sarma},
  Adam, Hwang, and Rossi}}]{dassarma2011}
\bibinfo{author}{\bibfnamefont{S.}~\bibnamefont{{Das Sarma}}},
  \bibinfo{author}{\bibfnamefont{S.}~\bibnamefont{Adam}},
  \bibinfo{author}{\bibfnamefont{E.~H.} \bibnamefont{Hwang}}, \bibnamefont{and}
  \bibinfo{author}{\bibfnamefont{E.}~\bibnamefont{Rossi}},
  \bibinfo{journal}{Rev. Mod. Phys.} \textbf{\bibinfo{volume}{83}},
  \bibinfo{pages}{407} (\bibinfo{year}{2011}).

\bibitem[{\citenamefont{Min et~al.}(2006)\citenamefont{Min, Hill, Sinitsyn,
  Sahu, Kleinman, and MacDonald}}]{min2006}
\bibinfo{author}{\bibfnamefont{H.}~\bibnamefont{Min}},
  \bibinfo{author}{\bibfnamefont{J.~E.} \bibnamefont{Hill}},
  \bibinfo{author}{\bibfnamefont{N.~A.} \bibnamefont{Sinitsyn}},
  \bibinfo{author}{\bibfnamefont{B.~R.} \bibnamefont{Sahu}},
  \bibinfo{author}{\bibfnamefont{L.}~\bibnamefont{Kleinman}}, \bibnamefont{and}
  \bibinfo{author}{\bibfnamefont{A.~H.} \bibnamefont{MacDonald}},
  \bibinfo{journal}{Phys. Rev. B} \textbf{\bibinfo{volume}{74}},
  \bibinfo{pages}{165310} (\bibinfo{year}{2006}).

\bibitem[{\citenamefont{Yao et~al.}(2007)\citenamefont{Yao, Ye, Qi, Zhang, and
  Fang}}]{yao2007}
\bibinfo{author}{\bibfnamefont{Y.~G.} \bibnamefont{Yao}},
  \bibinfo{author}{\bibfnamefont{F.}~\bibnamefont{Ye}},
  \bibinfo{author}{\bibfnamefont{X.~L.} \bibnamefont{Qi}},
  \bibinfo{author}{\bibfnamefont{S.~C.} \bibnamefont{Zhang}}, \bibnamefont{and}
  \bibinfo{author}{\bibfnamefont{Z.}~\bibnamefont{Fang}},
  \bibinfo{journal}{Phys. Rev. B} \textbf{\bibinfo{volume}{75}},
  \bibinfo{pages}{041401} (\bibinfo{year}{2007}).

\bibitem[{\citenamefont{Gmitra et~al.}(2009)\citenamefont{Gmitra, Konschuh,
  Ertler, Ambrosch-Draxl, and Fabian}}]{gmitra2009}
\bibinfo{author}{\bibfnamefont{M.}~\bibnamefont{Gmitra}},
  \bibinfo{author}{\bibfnamefont{S.}~\bibnamefont{Konschuh}},
  \bibinfo{author}{\bibfnamefont{C.}~\bibnamefont{Ertler}},
  \bibinfo{author}{\bibfnamefont{C.}~\bibnamefont{Ambrosch-Draxl}},
  \bibnamefont{and} \bibinfo{author}{\bibfnamefont{J.}~\bibnamefont{Fabian}},
  \bibinfo{journal}{Phys. Rev. B} \textbf{\bibinfo{volume}{80}},
  \bibinfo{pages}{235431} (\bibinfo{year}{2009}).

\bibitem[{\citenamefont{Rashba}(2009)}]{rashba2009}
\bibinfo{author}{\bibfnamefont{E.~I.} \bibnamefont{Rashba}},
  \bibinfo{journal}{Phys. Rev. B} \textbf{\bibinfo{volume}{79}},
  \bibinfo{pages}{161409} (\bibinfo{year}{2009}).

\bibitem[{\citenamefont{Dang et~al.}(2010)\citenamefont{Dang, Peng, Li, Wang,
  and Liu}}]{dang2010}
\bibinfo{author}{\bibfnamefont{W.~H.} \bibnamefont{Dang}},
  \bibinfo{author}{\bibfnamefont{H.~L.} \bibnamefont{Peng}},
  \bibinfo{author}{\bibfnamefont{H.}~\bibnamefont{Li}},
  \bibinfo{author}{\bibfnamefont{P.}~\bibnamefont{Wang}}, \bibnamefont{and}
  \bibinfo{author}{\bibfnamefont{Z.~F.} \bibnamefont{Liu}},
  \bibinfo{journal}{Nano Lett.} \textbf{\bibinfo{volume}{10}},
  \bibinfo{pages}{2870} (\bibinfo{year}{2010}).

\bibitem[{\citenamefont{{Song} et~al.}(2010)\citenamefont{{Song}, {Wang},
  {Jiang}, {Zhang}, {Chang}, {Wang}, {He}, {Chen}, {Jia}, {Wang}
  et~al.}}]{song2010}
\bibinfo{author}{\bibfnamefont{C.-L.} \bibnamefont{{Song}}},
  \bibinfo{author}{\bibfnamefont{Y.-L.} \bibnamefont{{Wang}}},
  \bibinfo{author}{\bibfnamefont{Y.-P.} \bibnamefont{{Jiang}}},
  \bibinfo{author}{\bibfnamefont{Y.}~\bibnamefont{{Zhang}}},
  \bibinfo{author}{\bibfnamefont{C.-Z.} \bibnamefont{{Chang}}},
  \bibinfo{author}{\bibfnamefont{L.}~\bibnamefont{{Wang}}},
  \bibinfo{author}{\bibfnamefont{K.}~\bibnamefont{{He}}},
  \bibinfo{author}{\bibfnamefont{X.}~\bibnamefont{{Chen}}},
  \bibinfo{author}{\bibfnamefont{J.-F.} \bibnamefont{{Jia}}},
  \bibinfo{author}{\bibfnamefont{Y.}~\bibnamefont{{Wang}}},
  \bibnamefont{et~al.}, \bibinfo{journal}{Applied Physics Letters}
  \textbf{\bibinfo{volume}{97}}, \bibinfo{pages}{143118}
  (\bibinfo{year}{2010}).

\bibitem[{\citenamefont{Weeks et~al.}(2011)\citenamefont{Weeks, Hu, Alicea,
  Franz, and Wu}}]{weeks2011}
\bibinfo{author}{\bibfnamefont{C.}~\bibnamefont{Weeks}},
  \bibinfo{author}{\bibfnamefont{J.}~\bibnamefont{Hu}},
  \bibinfo{author}{\bibfnamefont{J.}~\bibnamefont{Alicea}},
  \bibinfo{author}{\bibfnamefont{M.}~\bibnamefont{Franz}}, \bibnamefont{and}
  \bibinfo{author}{\bibfnamefont{R.}~\bibnamefont{Wu}}, \bibinfo{journal}{Phys.
  Rev. X} \textbf{\bibinfo{volume}{1}}, \bibinfo{pages}{021001}
  (\bibinfo{year}{2011}).

\bibitem[{\citenamefont{Jiang et~al.}(2012)\citenamefont{Jiang, Qiao, Liu, Shi,
  and Niu}}]{jiang2012}
\bibinfo{author}{\bibfnamefont{H.}~\bibnamefont{Jiang}},
  \bibinfo{author}{\bibfnamefont{Z.~H.} \bibnamefont{Qiao}},
  \bibinfo{author}{\bibfnamefont{H.~W.} \bibnamefont{Liu}},
  \bibinfo{author}{\bibfnamefont{J.~R.} \bibnamefont{Shi}}, \bibnamefont{and}
  \bibinfo{author}{\bibfnamefont{Q.}~\bibnamefont{Niu}},
  \bibinfo{journal}{Phys. Rev. Lett.} \textbf{\bibinfo{volume}{109}},
  \bibinfo{pages}{116803} (\bibinfo{year}{2012}).

\bibitem[{\citenamefont{Novoselov et~al.}(2005)\citenamefont{Novoselov, Jiang,
  Schedin, Booth, Khotkevich, Morozov, and Geim}}]{novoselov2005b}
\bibinfo{author}{\bibfnamefont{K.~S.} \bibnamefont{Novoselov}},
  \bibinfo{author}{\bibfnamefont{D.}~\bibnamefont{Jiang}},
  \bibinfo{author}{\bibfnamefont{F.}~\bibnamefont{Schedin}},
  \bibinfo{author}{\bibfnamefont{T.~J.} \bibnamefont{Booth}},
  \bibinfo{author}{\bibfnamefont{V.~V.} \bibnamefont{Khotkevich}},
  \bibinfo{author}{\bibfnamefont{S.~V.} \bibnamefont{Morozov}},
  \bibnamefont{and} \bibinfo{author}{\bibfnamefont{A.~K.} \bibnamefont{Geim}},
  \bibinfo{journal}{Proc. Nat. Acad. Sci. (USA)}
  \textbf{\bibinfo{volume}{102}}, \bibinfo{pages}{10451}
  (\bibinfo{year}{2005}).

\bibitem[{\citenamefont{Kim et~al.}(2009)\citenamefont{Kim, Nah, Jo,
  Shahrjerdi, Colombo, Yao, Tutuc, and Banerjee}}]{kim2009}
\bibinfo{author}{\bibfnamefont{S.}~\bibnamefont{Kim}},
  \bibinfo{author}{\bibfnamefont{J.}~\bibnamefont{Nah}},
  \bibinfo{author}{\bibfnamefont{I.}~\bibnamefont{Jo}},
  \bibinfo{author}{\bibfnamefont{D.}~\bibnamefont{Shahrjerdi}},
  \bibinfo{author}{\bibfnamefont{L.}~\bibnamefont{Colombo}},
  \bibinfo{author}{\bibfnamefont{Z.}~\bibnamefont{Yao}},
  \bibinfo{author}{\bibfnamefont{E.}~\bibnamefont{Tutuc}}, \bibnamefont{and}
  \bibinfo{author}{\bibfnamefont{S.~K.} \bibnamefont{Banerjee}},
  \bibinfo{journal}{Applied Physics Letters} \textbf{\bibinfo{volume}{94}},
  \bibinfo{eid}{062107} (\bibinfo{year}{2009}).

\bibitem[{\citenamefont{{Kim} et~al.}(2011)\citenamefont{{Kim}, {Jo}, {Nah},
  {Yao}, {Banerjee}, and {Tutuc}}}]{kim2010}
\bibinfo{author}{\bibfnamefont{S.}~\bibnamefont{{Kim}}},
  \bibinfo{author}{\bibfnamefont{I.}~\bibnamefont{{Jo}}},
  \bibinfo{author}{\bibfnamefont{J.}~\bibnamefont{{Nah}}},
  \bibinfo{author}{\bibfnamefont{Z.}~\bibnamefont{{Yao}}},
  \bibinfo{author}{\bibfnamefont{S.~K.} \bibnamefont{{Banerjee}}},
  \bibnamefont{and} \bibinfo{author}{\bibfnamefont{E.}~\bibnamefont{{Tutuc}}},
  \bibinfo{journal}{Phys. Rev. B} \textbf{\bibinfo{volume}{83}},
  \bibinfo{pages}{161401} (\bibinfo{year}{2011}).

\bibitem[{\citenamefont{{Ponomarenko} et~al.}(2011)\citenamefont{{Ponomarenko},
  {Geim}, {Zhukov}, {Jalil}, {Morozov}, {Novoselov}, {Grigorieva}, {Hill},
  {Cheianov}, {Fal'Ko} et~al.}}]{ponomarenko2011}
\bibinfo{author}{\bibfnamefont{L.~A.} \bibnamefont{{Ponomarenko}}},
  \bibinfo{author}{\bibfnamefont{A.~K.} \bibnamefont{{Geim}}},
  \bibinfo{author}{\bibfnamefont{A.~A.} \bibnamefont{{Zhukov}}},
  \bibinfo{author}{\bibfnamefont{R.}~\bibnamefont{{Jalil}}},
  \bibinfo{author}{\bibfnamefont{S.~V.} \bibnamefont{{Morozov}}},
  \bibinfo{author}{\bibfnamefont{K.~S.} \bibnamefont{{Novoselov}}},
  \bibinfo{author}{\bibfnamefont{I.~V.} \bibnamefont{{Grigorieva}}},
  \bibinfo{author}{\bibfnamefont{E.~H.} \bibnamefont{{Hill}}},
  \bibinfo{author}{\bibfnamefont{V.~V.} \bibnamefont{{Cheianov}}},
  \bibinfo{author}{\bibfnamefont{V.~I.} \bibnamefont{{Fal'Ko}}},
  \bibnamefont{et~al.}, \bibinfo{journal}{Nature Physics}
  \textbf{\bibinfo{volume}{7}}, \bibinfo{pages}{958} (\bibinfo{year}{2011}).

\bibitem[{\citenamefont{{Kim} and {Tutuc}}(2012)}]{kim2012}
\bibinfo{author}{\bibfnamefont{S.}~\bibnamefont{{Kim}}} \bibnamefont{and}
  \bibinfo{author}{\bibfnamefont{E.}~\bibnamefont{{Tutuc}}},
  \bibinfo{journal}{Solid State Communications} \textbf{\bibinfo{volume}{152}},
  \bibinfo{pages}{1283} (\bibinfo{year}{2012}).

\bibitem[{\citenamefont{{Gorbachev} et~al.}(2012)\citenamefont{{Gorbachev},
  {Geim}, {Katsnelson}, {Novoselov}, {Tudorovskiy}, {Grigorieva}, {MacDonald},
  {Morozov}, {Watanabe}, {Taniguchi} et~al.}}]{gorbachev2012}
\bibinfo{author}{\bibfnamefont{R.~V.} \bibnamefont{{Gorbachev}}},
  \bibinfo{author}{\bibfnamefont{A.~K.} \bibnamefont{{Geim}}},
  \bibinfo{author}{\bibfnamefont{M.~I.} \bibnamefont{{Katsnelson}}},
  \bibinfo{author}{\bibfnamefont{K.~S.} \bibnamefont{{Novoselov}}},
  \bibinfo{author}{\bibfnamefont{T.}~\bibnamefont{{Tudorovskiy}}},
  \bibinfo{author}{\bibfnamefont{I.~V.} \bibnamefont{{Grigorieva}}},
  \bibinfo{author}{\bibfnamefont{A.~H.} \bibnamefont{{MacDonald}}},
  \bibinfo{author}{\bibfnamefont{S.~V.} \bibnamefont{{Morozov}}},
  \bibinfo{author}{\bibfnamefont{K.}~\bibnamefont{{Watanabe}}},
  \bibinfo{author}{\bibfnamefont{T.}~\bibnamefont{{Taniguchi}}},
  \bibnamefont{et~al.}, \bibinfo{journal}{Nature Physics}
  \textbf{\bibinfo{volume}{8}}, \bibinfo{pages}{896} (\bibinfo{year}{2012}).

\bibitem[{\citenamefont{{Haigh} et~al.}(2012)\citenamefont{{Haigh}, {Gholinia},
  {Jalil}, {Romani}, {Britnell}, {Elias}, {Novoselov}, {Ponomarenko}, {Geim},
  and {Gorbachev}}}]{haigh2012}
\bibinfo{author}{\bibfnamefont{S.~J.} \bibnamefont{{Haigh}}},
  \bibinfo{author}{\bibfnamefont{A.}~\bibnamefont{{Gholinia}}},
  \bibinfo{author}{\bibfnamefont{R.}~\bibnamefont{{Jalil}}},
  \bibinfo{author}{\bibfnamefont{S.}~\bibnamefont{{Romani}}},
  \bibinfo{author}{\bibfnamefont{L.}~\bibnamefont{{Britnell}}},
  \bibinfo{author}{\bibfnamefont{D.~C.} \bibnamefont{{Elias}}},
  \bibinfo{author}{\bibfnamefont{K.~S.} \bibnamefont{{Novoselov}}},
  \bibinfo{author}{\bibfnamefont{L.~A.} \bibnamefont{{Ponomarenko}}},
  \bibinfo{author}{\bibfnamefont{A.~K.} \bibnamefont{{Geim}}},
  \bibnamefont{and}
  \bibinfo{author}{\bibfnamefont{R.}~\bibnamefont{{Gorbachev}}},
  \bibinfo{journal}{Nature Materials} \textbf{\bibinfo{volume}{11}},
  \bibinfo{pages}{764} (\bibinfo{year}{2012}).

\bibitem[{\citenamefont{Zhang and Rossi}(2013)}]{jzhang2013}
\bibinfo{author}{\bibfnamefont{J.}~\bibnamefont{Zhang}} \bibnamefont{and}
  \bibinfo{author}{\bibfnamefont{E.}~\bibnamefont{Rossi}},
  \bibinfo{journal}{Phys. Rev. Lett.} \textbf{\bibinfo{volume}{111}},
  \bibinfo{pages}{086804} (\bibinfo{year}{2013}).

\bibitem[{\citenamefont{Lopes~dos Santos et~al.}(2007)\citenamefont{Lopes~dos
  Santos, Peres, and Castro~Neto}}]{dossantos2007}
\bibinfo{author}{\bibfnamefont{J.~M.~B.} \bibnamefont{Lopes~dos Santos}},
  \bibinfo{author}{\bibfnamefont{N.~M.~R.} \bibnamefont{Peres}},
  \bibnamefont{and} \bibinfo{author}{\bibfnamefont{A.~H.}
  \bibnamefont{Castro~Neto}}, \bibinfo{journal}{Phys. Rev. Lett.}
  \textbf{\bibinfo{volume}{99}}, \bibinfo{pages}{256802}
  (\bibinfo{year}{2007}).

\bibitem[{\citenamefont{{Mele}}(2010)}]{mele2010}
\bibinfo{author}{\bibfnamefont{E.~J.} \bibnamefont{{Mele}}},
  \bibinfo{journal}{Phys. Rev. B} \textbf{\bibinfo{volume}{81}},
  \bibinfo{pages}{161405} (\bibinfo{year}{2010}).

\bibitem[{\citenamefont{Shallcross et~al.}(2010)\citenamefont{Shallcross,
  Sharma, Kandelaki, and Pankratov}}]{shallcross2010}
\bibinfo{author}{\bibfnamefont{S.~.} \bibnamefont{Shallcross}},
  \bibinfo{author}{\bibfnamefont{S.~.} \bibnamefont{Sharma}},
  \bibinfo{author}{\bibfnamefont{E.~.} \bibnamefont{Kandelaki}},
  \bibnamefont{and} \bibinfo{author}{\bibfnamefont{O.~A.}
  \bibnamefont{Pankratov}}, \bibinfo{journal}{Phys. Rev. B}
  \textbf{\bibinfo{volume}{81}}, \bibinfo{pages}{165105}
  (\bibinfo{year}{2010}).

\bibitem[{\citenamefont{Morell et~al.}(2010)\citenamefont{Morell, Correa,
  Vargas, Pacheco, and Barticevic}}]{suarez2010}
\bibinfo{author}{\bibfnamefont{E.~S.} \bibnamefont{Morell}},
  \bibinfo{author}{\bibfnamefont{J.~D.} \bibnamefont{Correa}},
  \bibinfo{author}{\bibfnamefont{P.}~\bibnamefont{Vargas}},
  \bibinfo{author}{\bibfnamefont{M.}~\bibnamefont{Pacheco}}, \bibnamefont{and}
  \bibinfo{author}{\bibfnamefont{Z.}~\bibnamefont{Barticevic}},
  \bibinfo{journal}{Phys. Rev. B} \textbf{\bibinfo{volume}{82}},
  \bibinfo{pages}{121407} (\bibinfo{year}{2010}).

\bibitem[{\citenamefont{Bistritzer and MacDonald}(2010)}]{bistritzer2010}
\bibinfo{author}{\bibfnamefont{R.}~\bibnamefont{Bistritzer}} \bibnamefont{and}
  \bibinfo{author}{\bibfnamefont{A.~H.} \bibnamefont{MacDonald}},
  \bibinfo{journal}{Phys. Rev. B} \textbf{\bibinfo{volume}{81}},
  \bibinfo{pages}{245412} (\bibinfo{year}{2010}).

\bibitem[{\citenamefont{Li et~al.}(2011)\citenamefont{Li, Martin, Buttiker, and
  Morpurgo}}]{li2010}
\bibinfo{author}{\bibfnamefont{J.}~\bibnamefont{Li}},
  \bibinfo{author}{\bibfnamefont{I.}~\bibnamefont{Martin}},
  \bibinfo{author}{\bibfnamefont{M.}~\bibnamefont{Buttiker}}, \bibnamefont{and}
  \bibinfo{author}{\bibfnamefont{A.~F.} \bibnamefont{Morpurgo}},
  \bibinfo{journal}{Nat. Phys.} \textbf{\bibinfo{volume}{7}},
  \bibinfo{pages}{38} (\bibinfo{year}{2011}).

\bibitem[{\citenamefont{Mele}(2011)}]{mele2011}
\bibinfo{author}{\bibfnamefont{E.~J.} \bibnamefont{Mele}},
  \bibinfo{journal}{Phys. Rev. B} \textbf{\bibinfo{volume}{84}},
  \bibinfo{pages}{235439} (\bibinfo{year}{2011}).

\bibitem[{\citenamefont{Bistritzer and
  MacDonald}(2011{\natexlab{a}})}]{bistritzer2011b}
\bibinfo{author}{\bibfnamefont{R.}~\bibnamefont{Bistritzer}} \bibnamefont{and}
  \bibinfo{author}{\bibfnamefont{A.~H.} \bibnamefont{MacDonald}},
  \bibinfo{journal}{Phys. Rev. B} \textbf{\bibinfo{volume}{84}},
  \bibinfo{pages}{035440} (\bibinfo{year}{2011}{\natexlab{a}}).

\bibitem[{\citenamefont{Bistritzer and
  MacDonald}(2011{\natexlab{b}})}]{bistritzer2011}
\bibinfo{author}{\bibfnamefont{R.}~\bibnamefont{Bistritzer}} \bibnamefont{and}
  \bibinfo{author}{\bibfnamefont{A.~H.} \bibnamefont{MacDonald}},
  \bibinfo{journal}{Proc. National Acad. Sciences United States Am.}
  \textbf{\bibinfo{volume}{108}}, \bibinfo{pages}{12233}
  (\bibinfo{year}{2011}{\natexlab{b}}).

\bibitem[{\citenamefont{Su\'arez~Morell
  et~al.}(2011)\citenamefont{Su\'arez~Morell, Vargas, Chico, and
  Brey}}]{morell2011}
\bibinfo{author}{\bibfnamefont{E.}~\bibnamefont{Su\'arez~Morell}},
  \bibinfo{author}{\bibfnamefont{P.}~\bibnamefont{Vargas}},
  \bibinfo{author}{\bibfnamefont{L.}~\bibnamefont{Chico}}, \bibnamefont{and}
  \bibinfo{author}{\bibfnamefont{L.}~\bibnamefont{Brey}},
  \bibinfo{journal}{Phys. Rev. B} \textbf{\bibinfo{volume}{84}},
  \bibinfo{pages}{195421} (\bibinfo{year}{2011}).

\bibitem[{\citenamefont{{Mele}}(2012)}]{mele2012}
\bibinfo{author}{\bibfnamefont{E.~J.} \bibnamefont{{Mele}}},
  \bibinfo{journal}{Journal of Physics D Applied Physics}
  \textbf{\bibinfo{volume}{45}}, \bibinfo{pages}{154004}
  (\bibinfo{year}{2012}).

\bibitem[{\citenamefont{{Kindermann} and {Mele}}(2011)}]{kindermann2011}
\bibinfo{author}{\bibfnamefont{M.}~\bibnamefont{{Kindermann}}}
  \bibnamefont{and} \bibinfo{author}{\bibfnamefont{E.~J.}
  \bibnamefont{{Mele}}}, \bibinfo{journal}{Phys. Rev. B}
  \textbf{\bibinfo{volume}{84}}, \bibinfo{pages}{161406}
  (\bibinfo{year}{2011}).

\bibitem[{\citenamefont{{Luican} et~al.}(2011)\citenamefont{{Luican}, {Li},
  {Reina}, {Kong}, {Nair}, {Novoselov}, {Geim}, and {Andrei}}}]{luican2011}
\bibinfo{author}{\bibfnamefont{A.}~\bibnamefont{{Luican}}},
  \bibinfo{author}{\bibfnamefont{G.}~\bibnamefont{{Li}}},
  \bibinfo{author}{\bibfnamefont{A.}~\bibnamefont{{Reina}}},
  \bibinfo{author}{\bibfnamefont{J.}~\bibnamefont{{Kong}}},
  \bibinfo{author}{\bibfnamefont{R.~R.} \bibnamefont{{Nair}}},
  \bibinfo{author}{\bibfnamefont{K.~S.} \bibnamefont{{Novoselov}}},
  \bibinfo{author}{\bibfnamefont{A.~K.} \bibnamefont{{Geim}}},
  \bibnamefont{and} \bibinfo{author}{\bibfnamefont{E.~Y.}
  \bibnamefont{{Andrei}}}, \bibinfo{journal}{Phys. Rev. Lett.}
  \textbf{\bibinfo{volume}{106}}, \bibinfo{pages}{126802}
  (\bibinfo{year}{2011}).

\bibitem[{\citenamefont{Yan et~al.}(2012)\citenamefont{Yan, Liu, Dou, Meng,
  Feng, Chu, Zhang, Liu, Nie, and He}}]{yan2012}
\bibinfo{author}{\bibfnamefont{W.}~\bibnamefont{Yan}},
  \bibinfo{author}{\bibfnamefont{M.}~\bibnamefont{Liu}},
  \bibinfo{author}{\bibfnamefont{R.-F.} \bibnamefont{Dou}},
  \bibinfo{author}{\bibfnamefont{L.}~\bibnamefont{Meng}},
  \bibinfo{author}{\bibfnamefont{L.}~\bibnamefont{Feng}},
  \bibinfo{author}{\bibfnamefont{Z.-D.} \bibnamefont{Chu}},
  \bibinfo{author}{\bibfnamefont{Y.}~\bibnamefont{Zhang}},
  \bibinfo{author}{\bibfnamefont{Z.}~\bibnamefont{Liu}},
  \bibinfo{author}{\bibfnamefont{J.-C.} \bibnamefont{Nie}}, \bibnamefont{and}
  \bibinfo{author}{\bibfnamefont{L.}~\bibnamefont{He}}, \bibinfo{journal}{Phys.
  Rev. Lett.} \textbf{\bibinfo{volume}{109}}, \bibinfo{pages}{126801}
  (\bibinfo{year}{2012}).

\bibitem[{\citenamefont{{Sanchez-Yamagishi}
  et~al.}(2012)\citenamefont{{Sanchez-Yamagishi}, {Taychatanapat}, {Watanabe},
  {Taniguchi}, {Yacoby}, and {Jarillo-Herrero}}}]{sanchez2012}
\bibinfo{author}{\bibfnamefont{J.~D.} \bibnamefont{{Sanchez-Yamagishi}}},
  \bibinfo{author}{\bibfnamefont{T.}~\bibnamefont{{Taychatanapat}}},
  \bibinfo{author}{\bibfnamefont{K.}~\bibnamefont{{Watanabe}}},
  \bibinfo{author}{\bibfnamefont{T.}~\bibnamefont{{Taniguchi}}},
  \bibinfo{author}{\bibfnamefont{A.}~\bibnamefont{{Yacoby}}}, \bibnamefont{and}
  \bibinfo{author}{\bibfnamefont{P.}~\bibnamefont{{Jarillo-Herrero}}},
  \bibinfo{journal}{Phys. Rev. Lett.} \textbf{\bibinfo{volume}{108}},
  \bibinfo{eid}{076601} (\bibinfo{year}{2012}).

\bibitem[{\citenamefont{{San-Jose} and {Prada}}(2013)}]{sanjose2013}
\bibinfo{author}{\bibfnamefont{P.}~\bibnamefont{{San-Jose}}} \bibnamefont{and}
  \bibinfo{author}{\bibfnamefont{E.}~\bibnamefont{{Prada}}},
  \bibinfo{journal}{Phys. Rev. B} \textbf{\bibinfo{volume}{88}},
  \bibinfo{pages}{121408(R)} (\bibinfo{year}{2013}).

\bibitem[{\citenamefont{Kumar and Nandkishore}(2013)}]{kumar2013}
\bibinfo{author}{\bibfnamefont{A.}~\bibnamefont{Kumar}} \bibnamefont{and}
  \bibinfo{author}{\bibfnamefont{R.}~\bibnamefont{Nandkishore}},
  \bibinfo{journal}{Phys. Rev. B} \textbf{\bibinfo{volume}{87}},
  \bibinfo{pages}{241108(R)} (\bibinfo{year}{2013}).

\bibitem[{\citenamefont{{Lu} and {Fertig}}(2014)}]{lu2014}
\bibinfo{author}{\bibfnamefont{C.-K.} \bibnamefont{{Lu}}} \bibnamefont{and}
  \bibinfo{author}{\bibfnamefont{H.~A.} \bibnamefont{{Fertig}}},
  \bibinfo{journal}{Phys. Rev. B} \textbf{\bibinfo{volume}{89}},
  \bibinfo{pages}{085408} (\bibinfo{year}{2014}).

\bibitem[{\citenamefont{Dean et~al.}(2010)\citenamefont{Dean, Young, Meric,
  Lee., Wang, Sorgenfrei, Watanabe, Taniguchi, Kim, Shepard et~al.}}]{dean2010}
\bibinfo{author}{\bibfnamefont{C.~R.} \bibnamefont{Dean}},
  \bibinfo{author}{\bibfnamefont{A.~F.} \bibnamefont{Young}},
  \bibinfo{author}{\bibfnamefont{I.}~\bibnamefont{Meric}},
  \bibinfo{author}{\bibfnamefont{C.}~\bibnamefont{Lee.}},
  \bibinfo{author}{\bibfnamefont{L.}~\bibnamefont{Wang}},
  \bibinfo{author}{\bibfnamefont{S.}~\bibnamefont{Sorgenfrei}},
  \bibinfo{author}{\bibfnamefont{K.}~\bibnamefont{Watanabe}},
  \bibinfo{author}{\bibfnamefont{T.}~\bibnamefont{Taniguchi}},
  \bibinfo{author}{\bibfnamefont{P.}~\bibnamefont{Kim}},
  \bibinfo{author}{\bibfnamefont{K.~L.} \bibnamefont{Shepard}},
  \bibnamefont{et~al.}, \bibinfo{journal}{Nature Nanotechnology}
  \textbf{\bibinfo{volume}{5}}, \bibinfo{pages}{726} (\bibinfo{year}{2010}).

\bibitem[{\citenamefont{{Xue} et~al.}(2011)\citenamefont{{Xue},
  {Sanchez-Yamagishi}, {Bulmash}, {Jacquod}, {Deshpande}, {Watanabe},
  {Taniguchi}, {Jarillo-Herrero}, and {Leroy}}}]{xue2011}
\bibinfo{author}{\bibfnamefont{J.}~\bibnamefont{{Xue}}},
  \bibinfo{author}{\bibfnamefont{J.}~\bibnamefont{{Sanchez-Yamagishi}}},
  \bibinfo{author}{\bibfnamefont{D.}~\bibnamefont{{Bulmash}}},
  \bibinfo{author}{\bibfnamefont{P.}~\bibnamefont{{Jacquod}}},
  \bibinfo{author}{\bibfnamefont{A.}~\bibnamefont{{Deshpande}}},
  \bibinfo{author}{\bibfnamefont{K.}~\bibnamefont{{Watanabe}}},
  \bibinfo{author}{\bibfnamefont{T.}~\bibnamefont{{Taniguchi}}},
  \bibinfo{author}{\bibfnamefont{P.}~\bibnamefont{{Jarillo-Herrero}}},
  \bibnamefont{and} \bibinfo{author}{\bibfnamefont{B.~J.}
  \bibnamefont{{Leroy}}}, \bibinfo{journal}{Nat. Mat.}
  \textbf{\bibinfo{volume}{10}}, \bibinfo{pages}{282} (\bibinfo{year}{2011}).

\bibitem[{\citenamefont{{Yankowitz} et~al.}(2012)\citenamefont{{Yankowitz},
  {Xue}, {Cormode}, {Sanchez-Yamagishi}, {Watanabe}, {Taniguchi},
  {Jarillo-Herrero}, {Jacquod}, and {Leroy}}}]{yankowitz2012}
\bibinfo{author}{\bibfnamefont{M.}~\bibnamefont{{Yankowitz}}},
  \bibinfo{author}{\bibfnamefont{J.}~\bibnamefont{{Xue}}},
  \bibinfo{author}{\bibfnamefont{D.}~\bibnamefont{{Cormode}}},
  \bibinfo{author}{\bibfnamefont{J.~D.} \bibnamefont{{Sanchez-Yamagishi}}},
  \bibinfo{author}{\bibfnamefont{K.}~\bibnamefont{{Watanabe}}},
  \bibinfo{author}{\bibfnamefont{T.}~\bibnamefont{{Taniguchi}}},
  \bibinfo{author}{\bibfnamefont{P.}~\bibnamefont{{Jarillo-Herrero}}},
  \bibinfo{author}{\bibfnamefont{P.}~\bibnamefont{{Jacquod}}},
  \bibnamefont{and} \bibinfo{author}{\bibfnamefont{B.~J.}
  \bibnamefont{{Leroy}}}, \bibinfo{journal}{Nature Physics}
  \textbf{\bibinfo{volume}{8}}, \bibinfo{pages}{382} (\bibinfo{year}{2012}).

\bibitem[{\citenamefont{{Ponomarenko} et~al.}(2013)\citenamefont{{Ponomarenko},
  {Gorbachev}, {Yu}, {Elias}, {Jalil}, {Patel}, {Mishchenko}, {Mayorov},
  {Woods}, {Wallbank} et~al.}}]{ponomarenko2013}
\bibinfo{author}{\bibfnamefont{L.~A.} \bibnamefont{{Ponomarenko}}},
  \bibinfo{author}{\bibfnamefont{R.~V.} \bibnamefont{{Gorbachev}}},
  \bibinfo{author}{\bibfnamefont{G.~L.} \bibnamefont{{Yu}}},
  \bibinfo{author}{\bibfnamefont{D.~C.} \bibnamefont{{Elias}}},
  \bibinfo{author}{\bibfnamefont{R.}~\bibnamefont{{Jalil}}},
  \bibinfo{author}{\bibfnamefont{A.~A.} \bibnamefont{{Patel}}},
  \bibinfo{author}{\bibfnamefont{A.}~\bibnamefont{{Mishchenko}}},
  \bibinfo{author}{\bibfnamefont{A.~S.} \bibnamefont{{Mayorov}}},
  \bibinfo{author}{\bibfnamefont{C.~R.} \bibnamefont{{Woods}}},
  \bibinfo{author}{\bibfnamefont{J.~R.} \bibnamefont{{Wallbank}}},
  \bibnamefont{et~al.}, \bibinfo{journal}{Nature}
  \textbf{\bibinfo{volume}{497}}, \bibinfo{pages}{594} (\bibinfo{year}{2013}).

\bibitem[{\citenamefont{Hunt et~al.}(2013)\citenamefont{Hunt,
  Sanchez-Yamagishi, Young, Yankowitz, LeRoy, Watanabe, Taniguchi, Moon,
  Koshino, Jarillo-Herrero et~al.}}]{hunt2013}
\bibinfo{author}{\bibfnamefont{B.}~\bibnamefont{Hunt}},
  \bibinfo{author}{\bibfnamefont{J.~D.} \bibnamefont{Sanchez-Yamagishi}},
  \bibinfo{author}{\bibfnamefont{A.~F.} \bibnamefont{Young}},
  \bibinfo{author}{\bibfnamefont{M.}~\bibnamefont{Yankowitz}},
  \bibinfo{author}{\bibfnamefont{B.~J.} \bibnamefont{LeRoy}},
  \bibinfo{author}{\bibfnamefont{K.}~\bibnamefont{Watanabe}},
  \bibinfo{author}{\bibfnamefont{T.}~\bibnamefont{Taniguchi}},
  \bibinfo{author}{\bibfnamefont{P.}~\bibnamefont{Moon}},
  \bibinfo{author}{\bibfnamefont{M.}~\bibnamefont{Koshino}},
  \bibinfo{author}{\bibfnamefont{P.}~\bibnamefont{Jarillo-Herrero}},
  \bibnamefont{et~al.}, \bibinfo{journal}{Science}
  \textbf{\bibinfo{volume}{340}}, \bibinfo{pages}{1427} (\bibinfo{year}{2013}).

\bibitem[{\citenamefont{Dean et~al.}(2013)\citenamefont{Dean, Wang, Maher,
  Forsythe, Ghahari, Gao, Katoch, Ishigami, Moon, Koshino et~al.}}]{dean2013}
\bibinfo{author}{\bibfnamefont{C.~R.} \bibnamefont{Dean}},
  \bibinfo{author}{\bibfnamefont{L.}~\bibnamefont{Wang}},
  \bibinfo{author}{\bibfnamefont{P.}~\bibnamefont{Maher}},
  \bibinfo{author}{\bibfnamefont{C.}~\bibnamefont{Forsythe}},
  \bibinfo{author}{\bibfnamefont{F.}~\bibnamefont{Ghahari}},
  \bibinfo{author}{\bibfnamefont{Y.}~\bibnamefont{Gao}},
  \bibinfo{author}{\bibfnamefont{J.}~\bibnamefont{Katoch}},
  \bibinfo{author}{\bibfnamefont{M.}~\bibnamefont{Ishigami}},
  \bibinfo{author}{\bibfnamefont{P.}~\bibnamefont{Moon}},
  \bibinfo{author}{\bibfnamefont{M.}~\bibnamefont{Koshino}},
  \bibnamefont{et~al.}, \bibinfo{journal}{Nature}
  \textbf{\bibinfo{volume}{497}}, \bibinfo{pages}{598} (\bibinfo{year}{2013}).

\bibitem[{\citenamefont{{Mucha-Kruczynski}
  et~al.}(2013)\citenamefont{{Mucha-Kruczynski}, {Wallbank}, and
  {Fal'ko}}}]{mucha2013}
\bibinfo{author}{\bibfnamefont{M.}~\bibnamefont{{Mucha-Kruczynski}}},
  \bibinfo{author}{\bibfnamefont{J.}~\bibnamefont{{Wallbank}}},
  \bibnamefont{and} \bibinfo{author}{\bibfnamefont{V.~I.}
  \bibnamefont{{Fal'ko}}}, \bibinfo{journal}{Phys. Rev. B}
  \textbf{\bibinfo{volume}{88}}, \bibinfo{pages}{205418}
  (\bibinfo{year}{2013}).

\bibitem[{\citenamefont{Jin and Jhi}(2013)}]{jin2013}
\bibinfo{author}{\bibfnamefont{K.-H.} \bibnamefont{Jin}} \bibnamefont{and}
  \bibinfo{author}{\bibfnamefont{S.-H.} \bibnamefont{Jhi}},
  \bibinfo{journal}{Phys. Rev. B} \textbf{\bibinfo{volume}{87}},
  \bibinfo{pages}{075442} (\bibinfo{year}{2013}).

\bibitem[{\citenamefont{Zhang et~al.}(2009)\citenamefont{Zhang, Liu, Qi, Dai,
  Fang, and Zhang}}]{zhang2009ti}
\bibinfo{author}{\bibfnamefont{H.~J.} \bibnamefont{Zhang}},
  \bibinfo{author}{\bibfnamefont{C.~X.} \bibnamefont{Liu}},
  \bibinfo{author}{\bibfnamefont{X.~L.} \bibnamefont{Qi}},
  \bibinfo{author}{\bibfnamefont{X.}~\bibnamefont{Dai}},
  \bibinfo{author}{\bibfnamefont{Z.}~\bibnamefont{Fang}}, \bibnamefont{and}
  \bibinfo{author}{\bibfnamefont{S.~C.} \bibnamefont{Zhang}},
  \bibinfo{journal}{Nature Phys.} \textbf{\bibinfo{volume}{5}},
  \bibinfo{pages}{438} (\bibinfo{year}{2009}).

\bibitem[{\citenamefont{Liu et~al.}(2010)\citenamefont{Liu, Qi, Zhang, Dai,
  Fang, and Zhang}}]{liucx2010}
\bibinfo{author}{\bibfnamefont{C.-X.} \bibnamefont{Liu}},
  \bibinfo{author}{\bibfnamefont{X.-L.} \bibnamefont{Qi}},
  \bibinfo{author}{\bibfnamefont{H.~J.} \bibnamefont{Zhang}},
  \bibinfo{author}{\bibfnamefont{X.}~\bibnamefont{Dai}},
  \bibinfo{author}{\bibfnamefont{Z.}~\bibnamefont{Fang}}, \bibnamefont{and}
  \bibinfo{author}{\bibfnamefont{S.-C.} \bibnamefont{Zhang}},
  \bibinfo{journal}{Phys. Rev. B} \textbf{\bibinfo{volume}{82}},
  \bibinfo{pages}{045122} (\bibinfo{year}{2010}).

\bibitem[{\citenamefont{Fu}(2009)}]{fu2009b}
\bibinfo{author}{\bibfnamefont{L.}~\bibnamefont{Fu}}, \bibinfo{journal}{Phys.
  Rev. Lett.} \textbf{\bibinfo{volume}{103}}, \bibinfo{pages}{266801}
  (\bibinfo{year}{2009}).

\bibitem[{\citenamefont{Chen et~al.}(2009)\citenamefont{Chen, Cullen, Jang,
  Fuhrer, and Williams}}]{chen2009}
\bibinfo{author}{\bibfnamefont{J.-H.} \bibnamefont{Chen}},
  \bibinfo{author}{\bibfnamefont{W.~G.} \bibnamefont{Cullen}},
  \bibinfo{author}{\bibfnamefont{C.}~\bibnamefont{Jang}},
  \bibinfo{author}{\bibfnamefont{M.~S.} \bibnamefont{Fuhrer}},
  \bibnamefont{and} \bibinfo{author}{\bibfnamefont{E.~D.}
  \bibnamefont{Williams}}, \bibinfo{journal}{Phys. Rev. Lett.}
  \textbf{\bibinfo{volume}{102}}, \bibinfo{pages}{236805}
  (\bibinfo{year}{2009}).

\bibitem[{\citenamefont{Analytis et~al.}(2010)\citenamefont{Analytis, Chu,
  Chen, Corredor, McDonald, Shen, and Fisher}}]{analytis2010}
\bibinfo{author}{\bibfnamefont{J.~G.} \bibnamefont{Analytis}},
  \bibinfo{author}{\bibfnamefont{J.~H.} \bibnamefont{Chu}},
  \bibinfo{author}{\bibfnamefont{Y.~L.} \bibnamefont{Chen}},
  \bibinfo{author}{\bibfnamefont{F.}~\bibnamefont{Corredor}},
  \bibinfo{author}{\bibfnamefont{R.~D.} \bibnamefont{McDonald}},
  \bibinfo{author}{\bibfnamefont{Z.~X.} \bibnamefont{Shen}}, \bibnamefont{and}
  \bibinfo{author}{\bibfnamefont{I.~R.} \bibnamefont{Fisher}},
  \bibinfo{journal}{Phys. Rev. B} \textbf{\bibinfo{volume}{81}},
  \bibinfo{pages}{205407} (\bibinfo{year}{2010}).

\bibitem[{\citenamefont{Hsieh et~al.}(2009)\citenamefont{Hsieh, Xia, Qian,
  Wray, Meier, Dil, Osterwalder, Patthey, Fedorov, Lin et~al.}}]{hsieh2009}
\bibinfo{author}{\bibfnamefont{D.}~\bibnamefont{Hsieh}},
  \bibinfo{author}{\bibfnamefont{Y.}~\bibnamefont{Xia}},
  \bibinfo{author}{\bibfnamefont{D.}~\bibnamefont{Qian}},
  \bibinfo{author}{\bibfnamefont{L.}~\bibnamefont{Wray}},
  \bibinfo{author}{\bibfnamefont{F.}~\bibnamefont{Meier}},
  \bibinfo{author}{\bibfnamefont{J.~H.} \bibnamefont{Dil}},
  \bibinfo{author}{\bibfnamefont{J.}~\bibnamefont{Osterwalder}},
  \bibinfo{author}{\bibfnamefont{L.}~\bibnamefont{Patthey}},
  \bibinfo{author}{\bibfnamefont{A.~V.} \bibnamefont{Fedorov}},
  \bibinfo{author}{\bibfnamefont{H.}~\bibnamefont{Lin}}, \bibnamefont{et~al.},
  \bibinfo{journal}{Phys. Rev. Lett.} \textbf{\bibinfo{volume}{103}},
  \bibinfo{pages}{146401} (\bibinfo{year}{2009}).

\bibitem[{\citenamefont{{Steinberg} et~al.}(2010)\citenamefont{{Steinberg},
  {Gardner}, {Lee}, and {Jarillo-Herrero}}}]{steinberg2010}
\bibinfo{author}{\bibfnamefont{H.}~\bibnamefont{{Steinberg}}},
  \bibinfo{author}{\bibfnamefont{D.~R.} \bibnamefont{{Gardner}}},
  \bibinfo{author}{\bibfnamefont{Y.~S.} \bibnamefont{{Lee}}}, \bibnamefont{and}
  \bibinfo{author}{\bibfnamefont{P.}~\bibnamefont{{Jarillo-Herrero}}},
  \bibinfo{journal}{Nano Letters} \textbf{\bibinfo{volume}{10}},
  \bibinfo{pages}{5032} (\bibinfo{year}{2010}).

\bibitem[{\citenamefont{Checkelsky et~al.}(2011)\citenamefont{Checkelsky, Hor,
  Cava, and Ong}}]{checkelsky2011}
\bibinfo{author}{\bibfnamefont{J.~G.} \bibnamefont{Checkelsky}},
  \bibinfo{author}{\bibfnamefont{Y.~S.} \bibnamefont{Hor}},
  \bibinfo{author}{\bibfnamefont{R.~J.} \bibnamefont{Cava}}, \bibnamefont{and}
  \bibinfo{author}{\bibfnamefont{N.~P.} \bibnamefont{Ong}},
  \bibinfo{journal}{Phys. Rev. Lett.} \textbf{\bibinfo{volume}{106}},
  \bibinfo{pages}{196801} (\bibinfo{year}{2011}).

\bibitem[{\citenamefont{Zhang et~al.}(2010)\citenamefont{Zhang, He, Chang,
  Song, Wang, Chen, Jia, Fang, Dai, Shan et~al.}}]{zhangy2010}
\bibinfo{author}{\bibfnamefont{Y.}~\bibnamefont{Zhang}},
  \bibinfo{author}{\bibfnamefont{K.}~\bibnamefont{He}},
  \bibinfo{author}{\bibfnamefont{C.-Z.} \bibnamefont{Chang}},
  \bibinfo{author}{\bibfnamefont{C.-L.} \bibnamefont{Song}},
  \bibinfo{author}{\bibfnamefont{L.-L.} \bibnamefont{Wang}},
  \bibinfo{author}{\bibfnamefont{X.}~\bibnamefont{Chen}},
  \bibinfo{author}{\bibfnamefont{J.-F.} \bibnamefont{Jia}},
  \bibinfo{author}{\bibfnamefont{Z.}~\bibnamefont{Fang}},
  \bibinfo{author}{\bibfnamefont{X.}~\bibnamefont{Dai}},
  \bibinfo{author}{\bibfnamefont{W.-Y.} \bibnamefont{Shan}},
  \bibnamefont{et~al.}, \bibinfo{journal}{Nature Physics}
  \textbf{\bibinfo{volume}{6}}, \bibinfo{pages}{584} (\bibinfo{year}{2010}).

\bibitem[{\citenamefont{Taskin et~al.}(2012)\citenamefont{Taskin, Sasaki,
  Segawa, and Ando}}]{taskin2012}
\bibinfo{author}{\bibfnamefont{A.~A.} \bibnamefont{Taskin}},
  \bibinfo{author}{\bibfnamefont{S.}~\bibnamefont{Sasaki}},
  \bibinfo{author}{\bibfnamefont{K.}~\bibnamefont{Segawa}}, \bibnamefont{and}
  \bibinfo{author}{\bibfnamefont{Y.}~\bibnamefont{Ando}},
  \bibinfo{journal}{Phys. Rev. Lett.} \textbf{\bibinfo{volume}{109}},
  \bibinfo{pages}{066803} (\bibinfo{year}{2012}).

\bibitem[{\citenamefont{Hutasoit and Stanescu}(2011)}]{hutasoit2011}
\bibinfo{author}{\bibfnamefont{J.~A.} \bibnamefont{Hutasoit}} \bibnamefont{and}
  \bibinfo{author}{\bibfnamefont{T.~D.} \bibnamefont{Stanescu}},
  \bibinfo{journal}{Phys. Rev. B} \textbf{\bibinfo{volume}{84}},
  \bibinfo{pages}{085103} (\bibinfo{year}{2011}).

\bibitem[{not()}]{note2013}
\bibinfo{note}{Note that the Dirac Hamiltonian of graphene in the sublattice
  basis has a phase factor on the off-diagonal terms to take into account a
  rotational transformation of coordinates due to the choice of the current BZ
  orientation.}

\end{thebibliography}



\end{document}